\documentclass{article}
\usepackage{amsmath,amssymb,mathtools,cite,graphicx,fancyhdr}
\usepackage{bbold,color,mathpazo}
\usepackage{caption}
\begin{document}
\lhead{Wronskian differential formula for $k$-confluent SUSY QM}
\rhead{D. Bermudez}

\title{Wronskian differential formula for $k$-confluent SUSY QM}
\author{David Bermudez \\
{\sl Departamento de F\'{\i}sica, Cinvestav, A.P. 14-740, 07000 M\'exico D.F., Mexico}\\
{\it email:} dbermudez@fis.cinvestav.mx}

\date{}

\maketitle

\begin{abstract}
The confluent SUSY QM usually involves a second-order SUSY transformation where the two factorization energies converge to a single value. In order to achieve it, one generally needs to solve an indefinite integral, which limits the actual systems to which it can be applied. Nevertheless, not so long ago, an alternative method to achieve this transformation was developed through a Wronskian differential formula [Phys Lett. A 3756 (2012) 692]. In the present work, we consider the $k$-confluent SUSY transformation, where $k$ factorization energies merge into a single value, and we develop a generalized Wronskian differential formula for this case. Furthermore, we explicitly work out general formulas for the third- and fourth-order cases and we present as examples the free particle and the single-gap Lam\'e potentials.\\

\end{abstract}

\section{Introduction}\label{intro}

\subsection{Background}\label{back}
The supersymmetric quantum mechanics (SUSY QM), factorization method or intertwining technique are equivalent algebraic methods used in quantum mechanics to generate new solvable potentials. This technique starts from known solutions (either particular or general) of an eigenvalue problem in quantum mechanics and develop a deformation of it such that the solution of a new problem is given in terms of the original solution plus some parameters from the transformation.

In this way we are able to obtain parametric families of new potentials, from which usually one tries to find those without added singularities in order not to change the domain of definition and to obtain physically acceptable quantum systems, which otherwise would be just mathematically acceptable. After years of development, this method has become highly technical, with a lot of options such as: increasing the order (from first to $k$th), changing the type of transformation (real, confluent and complex), among others. For a review of this topic see Refs. \cite{MR04,Fer10} and references there in.

In this work we deal with a specific case of higher-order SUSY transformations, the confluent case, in which several factorization energies converge to the same value \cite{FS03,FS05,BFF12}. Taking such limit appropriately, this transformation leads to more flexibility for spectral design compared to the usual real case. Let us note that the standard method to perform this transformation requires to solve an indefinite integral, which is sometimes difficult to accomplish. Despite that, there has been a recent interest in this type of transformations, both in the study of its general properties \cite{Sch13,CS15,CS15b,GQ15} and in its applications \cite{BFF12,BF13b,CS14,AP15,CJP15,FH15,SW15}.

In recent times \cite{BFF12}, an alternative differential method has been developed in terms of Wronskians, which allows us to perform a confluent transformation without the need of solving integrals and, therefore, expanding the pool of potentials we can deal with. In the present work, we will see that this differential method is more general than it was previously realized. Here we will prove the general algorithm for the $k$th-order confluent SUSY QM or $k$-confluent SUSY QM \cite{Sch13,GQ15} (note that for $k=3$ it has been called {\it hyperconfluent} SUSY QM \cite{FS11}). Furthermore, we will explicitly develop two different ways to calculate both the 3- and the 4-confluent SUSY transformation.

Finally, we will apply these new formulas to two interesting systems. The first one is the free particle, whose 3-confluent SUSY transformation is already known and it can help us to check our results. The second is the single-gap Lam\'e potential, whose higher-order confluent SUSY transformation has not been addressed before.

\subsection{$k$th-order SUSY QM}\label{susy}
In the $k$th-order SUSY QM \cite{Fer10}, one starts from the following intertwining relationship,
\begin{equation}
	H_k B_k^+=B_k^+H_0,
\end{equation}
where $H_0$ is the original Hamiltonian and $H_k$ is the new one, still to be found. Usually $H_0$ is exactly-solvable but in general whatever we know (or do not know) about this system will be inherited by the new one. Furthermore, the more we know about $H_0$, the more flexibility we have to perform a SUSY transformation. Both Hamiltonians define their respective potentials $V_0$, $V_k$ and $B_k$ is a general $k$th-order differential intertwining operator, all of them given by:
\begin{subequations}
	\begin{align}
		H_0 &=-\partial_x^2 + V_0(x),\\
		H_k &= - \partial_x^2 + V_k(x),\\
		B_k^+ &= \sum_{j=0}^k (-1)^j b_{j}(x)\partial_x^j\, ,
	\end{align}
\end{subequations}
where the $b_{j}(x)$ are functions to be found. Also $B_k^+$ and $B_k^-=(B_k^+)^\dag$ fulfill the following factorizations
\begin{subequations}
	\begin{align}
		B_k^- B_k^+ &= \prod_{j=1}^k (H_0 - \epsilon_j),\\
		B_k^+ B_k^- &= \prod_{j=1}^k (H_k - \epsilon_j),
	\end{align}\label{facs}
\end{subequations}
with $\epsilon_j$ being the factorization energies for the transformation functions $u_j$.

According to the usual SUSY QM, the new potential $V_k$ will be a deformation of the initial one $V_0$ of the form \cite{Fer10,Sch13,Cru55,Ber13}:
\begin{equation}\label{hosusy}
	V_k(x) = V_0(x)-2\,\partial_x^2[ \ln W(u_1,u_2,\dots ,u_k)],
\end{equation}
where the functions $u_j$ are $k$ {\it mathematical} solutions of the original Schr\"odinger equation for $V_0$ and $W$ is the Wronskian. We use the adjective {\it mathematical} to refer to the solutions of the eigenvalue problem without the corresponding boundary conditions, which are the \textit{physical} ones. In this work we will fix a specific domain only in the applications part, such that the theory will be as general as possible.

From now on we will consider the special case of the SUSY transformation where all the factorization energies $\epsilon_j$ in Eqs.~\eqref{facs} converge to the same value, i.e., $\epsilon_j=\epsilon$ for $j=1,\dots , k$. In this case, Eq.~\eqref{hosusy} still remains valid (see Theorem 1 in Ref. \cite{Sch13}) but the $u_j$ are now generalized eigenfunctions associated with the same eigenvalue $\epsilon$. In fact, usually obtaining a generalized eigenvector of higher rank is quite difficult but, as we will see, departing from a known eigenvector there is a simple way to solve all the generalized eigenvalue equations associated with the same eigenvalue \cite{FS05,DK67,HJM06} and to find the related Jordan block (also known as Jordan chain or Jordan cycle).

Usually in a $k$-SUSY transformation, up to $k$ new levels could be added. These states could be {\it mathematical} or {\it physical} according to their boundary conditions in the sense we have just described and only the ones that are physical are counted as added levels to the spectrum of the new system. Nevertheless, due to the nature of the confluent case, where all the transformation energies converge to the same value, only one level could potentially be added, depending again if the state is physical or not. The new eigenstate is given by \cite{Sch13,FS11}:

\begin{equation}
	\psi_k (x,\epsilon)=\frac{W(u_1,u_2,\dots , u_{k-1})}{W(u_1,u_2,\dots , u_{k})},\label{hopsi}
\end{equation}
where $k\geq 2$ in order to have at least two factorization energies converging and $W(u)=u$ and again in the confluent case these functions $u_j$ are generalized eigenfunctions.

In the usual confluent SUSY transformations, this system is solved through indefinite integrals and from there a new family of potentials can be derived. In this work we will look for the general solution of Eq.~\eqref{hosusy} in a different way, through Wronskians and derivatives.

\subsection{Generalized eigenvectors and Jordan blocks}\label{eigen}
The equation that defines a usual eigenvector $\psi$ can be written as
\begin{subequations}
	\begin{align}
		(H-E)\psi & = 0,\\
		\psi & \neq 0,
	\end{align}
\end{subequations}
where $H$ is any operator and $E$ is its eigenvalue.

There is also a generalized definition given by
\begin{subequations}
	\begin{align}
		(H-E)^k \psi & = 0,\\
		(H-E)^{k-1} \psi & \neq 0,
	\end{align}
\end{subequations}
where $\psi$ is known as a {\it generalized eigenvector of rank $k$} of the operator $H$ and $E$ is the {\it generalized eigenvalue} of $H$ \cite{DK67}.

Starting from a standard eigenvector (of rank 1) we can form a chain of consecutive solutions for the generalized eigenvectors of rank $k$. In the simplest case, let $\psi_1$ be an eigenvector of rank 1 of $H$, i.e., $(H-E)\psi_1 = 0$, then an eigenvector of second-rank must fulfill $(H-E)^2\psi_2 =0$, which can be accomplished if we ask that $(H-E)\psi_2 =\psi_1$. Therefore we get the following system of equations:
\begin{subequations}
	\begin{align}
		(H-E)\psi_1 & = 0,\\
		(H-E)\psi_2 & = \psi_1,
	\end{align}
\end{subequations}
which is known as a Jordan block, Jordan chain or Jordan cycle, since the matrix representation of $H$ would be in this case
\begin{equation}
	(H)= \left( \begin{array}{cc}
		E & 1  \\
		0 & E
	\end{array}\right),
\end{equation}
i.e., a Jordan block. Generalized eigenvectors and their corresponding Jordan blocks of higher-rank can be obtained similarly and the matrix representation would be a $k\times k$ matrix with the eigenvalue $E$ in the diagonal and $1$ in the superdiagonal. Our final purpose here is to study the Jordan block of rank $k$; in order to accomplish this, let us first study some low-order cycles and then address the $k$th-order one.

In the rest of the paper we will derive the differential Wronskian formulas for confluent SUSY transformations of orders 2, 3, 4 and $k$. Each of these cases will be developed in three steps: first, we will define the Jordan block of the corresponding length. Then, we will obtain the general solution of the generalized eigenvectors for the corresponding rank and from them we will finally obtain the differential formula for the $k$-confluent SUSY transformation. In Section~\ref{apps} we will apply these formulas to two examples: the free particle and the single-gap Lam\'e potentials. We will finish this work with our Conclusions and an Appendix.

\section{Differential formula for $2$-confluent}\label{second}

\subsection{Jordan block of length 2}\label{2jordan}
Let us consider the following pair of generalized eigenfunctions of $H$, of first ($u_1$) and second ($u_2$) rank, associated with the eigenvalue $\epsilon$:
\begin{subequations}
	\begin{align}
		(H-\epsilon)u_1 & = 0,\label{uv1}\\
		(H-\epsilon)u_2 & = u_1, \label{uv2}
	\end{align}\label{jordan2}
\end{subequations}
\hspace{-2.7mm} defining a Jordan block of length 2. The functions $u_j$ do not necessarily have a physical interpretation, i.e., they could be {\it mathematical} eigenfunctions of $H$. Eq.~\eqref{uv1} is a second-order homogeneous equation, therefore its basis has two linearly independent solutions. They can be taken as $u$ and its orthogonal function $v\equiv u^\perp$ defined by $W(u,v) = 1$ (this is for simplicity, indeed the only condition is that $W(u,v) \neq 0$). Without lost of generality, we can choose 
\begin{equation}
	u_1=u,\label{u1}
\end{equation}
and then
\begin{equation}
	(H-\epsilon)u=0.\label{schu}
\end{equation}
From $W(u,v) = 1$ we can obtain that
\begin{equation}\label{defv}
	v(x) = u(x)\int\frac{\text{d}x}{u^2(x)}.
\end{equation}
We can easily prove that $v(x)$ also solves the Schr\"odinger equation~\eqref{schu} by applying $\partial_x^2$ to Eq.~\eqref{defv}, i.e.,
\begin{equation}
	(H-\epsilon)v=0.\label{schv}
\end{equation}

\subsection{Eigenvectors of rank 2}\label{2eigen}
The second equation of the Jordan block \eqref{uv2} is an inhomogeneous second-order differential equation and its general solution takes the following form
\begin{equation}
	u_2=u_2^{(h)}+u_2^{(p)},
\end{equation}
where $u_2^{(h)}$ is the general solution of the homogeneous equation and $u_2^{(p)}$ denotes a particular solution of the inhomogeneous one. We can use the same basis to describe the homogeneous solution of this second equation because we are dealing with generalized eigenvectors of the same eigenvalue \cite{DK67}. Then, the homogeneous solution is given by a linear combination of our basis functions $u,v$,
\begin{equation}
	u_2^{(h)}=C_1 u+D_1 v,
\end{equation}
with $C_1,\,D_1 \in \mathbb{R}$.

In order to find the particular solution $u_2^{(p)}$, let us suppose from now on that $u$ and its parametric derivative with respect to $\epsilon$, $\partial_\epsilon u$, are well defined continuous functions in a neighborhood of $\epsilon$. Hence, by deriving Eq.~\eqref{schu} with respect to $\epsilon$ we obtain
\begin{equation}
	\left(H-\epsilon\right)\partial_\epsilon u = u, \label{para1}
\end{equation}
where we assume that $u=u(x,\epsilon)$, $\partial_\epsilon \partial_x u=\partial_x \partial_\epsilon u$ and $H=H(x)\neq H(\epsilon)$. If we compare Eqs. \eqref{uv2} and \eqref{para1}, we can see that
\begin{equation}
	u_2^{(p)} = \partial_\epsilon u,
\end{equation}
i.e., $\partial_\epsilon u$ is a particular solution of the inhomogeneous equation we were looking for. Finally, the general solution of Eq.~\eqref{uv2} is given by
\begin{equation}
	u_2= C_1 u + D_1 v + \partial_\epsilon u .\label{u2}
\end{equation}
This is is the most general solution that closes the Jordan block of length 2. The apparent asymmetry between $u$ and $v$ due to the last term of this equation is produced by the initial choice of $u_1$ as $u$. Nevertheless, there is no loss of generality since this is just a parameter choice that simplifies the equations by eliminating one term at the end (which for higher-orders will be quite useful).

\subsection{Formula for $2$-confluent SUSY}\label{2formula}
We can obtain a new Hamiltonian from Eq.~\eqref{hosusy} by calculating the Wronskian of the two solutions $u_1,u_2$ of the Jordan block. This is straightforward, and the result is (it has 2! terms):
\begin{equation}
	W(u_1,u_2)= D_1 + W\left(u,\partial_\epsilon u\right), \label{wronsk2}
\end{equation}
where we make use of $W(u,v)=1$. Thus, the new potential $V_2(x)$ from the higher-order SUSY formula \eqref{hosusy} becomes
\begin{equation}
	V_2(x)=V(x)-2\, \partial_x^2 \ln \left[D_1 + W\left(u,\partial_\epsilon u\right)\right], \label{formula2}
\end{equation}
which represents an alternative way to calculate $V_2(x)$ through the confluent second-order SUSY transformation. Furthermore, the new eigenstate is given by Eq.~\eqref{hopsi} as
\begin{equation}
	\psi_2 (x,\epsilon)=\frac{u}{D_1 + W\left(u,\partial_\epsilon u\right)}.\label{psi2}
\end{equation}
Note that these equations were first introduced by Bermudez et al. in Ref.~\cite{BFF12}, where the general 2-confluent case for the free particle and single-gap Lam\'e potentials were also solved. Moreover, special cases for the free particle were previously addressed by Matveev \cite{Mat92} and Stahlhofen \cite{Sta95} without the constant $D_1$. We must remark that this constant is very important, as it characterizes a one-parametric family of SUSY partner potentials for each factorization energy $\epsilon$ from which we can choose the ones without singularities, i.e., the meaningful quantum systems. Due to the Wronskian structure and our choice of $u_1=u$, the constant $C_1$ does not appear in the Wronskian formula in Eq.~\eqref{wronsk2}, therefore in the end we obtain a 2-parametric family of potentials, with $\epsilon ,D_1$ as parameters.

\section{Differential formula for $3$-confluent}\label{third}

\subsection{Jordan block of length 3}\label{3jordan}
The Jordan block of length 3 is
\begin{subequations}
	\begin{align}
		(H-\epsilon)u_1 &=0,\\
		(H-\epsilon)u_2 &=u_1,\\
		(H-\epsilon)u_3 &=u_2.\label{jordan3c}
	\end{align}\label{jordan3}
\end{subequations}
\hspace{-1mm}In order to solve it, we can follow an analogous procedure; thus, $u_1$ and $u_2$ are also those given by Eqs. \eqref{u1} and \eqref{u2} because they close the Jordan block of length 2, we only need to solve Eq. {\eqref{jordan3c}, which we will do next.

\subsection{Eigenvectors of rank 3}\label{3eigen}

The eigenvector $u_3$ of rank 3 is also given as a linear combination of a homogeneous and an inhomogeneous part
\begin{equation}
	u_3=u_3^{(h)}+u_3^{(p)},
\end{equation}
The general solution of the homogeneous part can be expressed in the basis $u,v$ because $u_3$ is also associated with the same eigenvalue $\epsilon$; therefore:
\begin{equation}
	u_3^{(h)}=C_2u+D_2v,
\end{equation}
where $C_2, D_2 \in \mathbb{R}$ are new constants. A particular solution $u_3^{(p)}$ can be found if we apply the operator $\partial_\epsilon$ to both sides of Eq.~\eqref{para1}, leading to:
\begin{equation}
	(H-\epsilon)\frac{\partial_\epsilon^2 u}{2}=\partial_\epsilon u.\label{para2}
\end{equation}
Then, replacing Eqs.\eqref{para1} and \eqref{para2} into Eq.~\eqref{jordan3c}, we obtain:
\begin{equation}
	u_3^{(p)}=C_1\partial_\epsilon u +D_1\partial_\epsilon v+\frac{\partial_\epsilon^2 u}{2},
\end{equation}
and therefore the general solution $u_3$ is given by
\begin{equation}
	u_3=C_2u+D_2v+C_1\partial_\epsilon u +D_1\partial_\epsilon v+\frac{\partial_\epsilon^2 u}{2},\label{u3}
\end{equation}
where $C_1,D_1$ are the same as in Eq.~\eqref{u2}.

\subsection{Formula for 3-confluent SUSY}\label{3formula}
The three functions $u_1,u_2,u_3$ that close the Jordan block of length 3 are given by Eqs. \eqref{u1}, \eqref{u2} and \eqref{u3}. If we calculate their Wronskian we obtain (it should contain $3!$ terms):
\begin{align}\label{wronsk31}
	W(u_1,u_2,u_3)=&(C_1D_1-D_2)W(u,v,\partial_\epsilon u)+D_1^2W(u,v,\partial_\epsilon v)+D_1W(u,\partial_\epsilon u, \partial_\epsilon v)\nonumber\\
	&+\frac{D_1}{2}W(u,v,\partial_\epsilon^2 u)+\frac{1}{2}W(u,\partial_\epsilon u,\partial_\epsilon^2 u).
\end{align}
We can further simplify this equation by using the Wronskian identities that are found in the Appendix. In particular, we use Eq.~\eqref{app3} to express the three-function Wronskians in terms of two-function ones, and we obtain
\begin{align}\label{wronsk32}
	W(u_1,u_2,u_3)=&\left[D_2-C_1D_1+D_1\, W(u,\partial_\epsilon v) +\frac{1}{2}W(u,\partial_\epsilon^2 u)\right]u-D_1^2v-D_1\partial_\epsilon u \nonumber\\
	&-(D_1v+\partial_\epsilon u)W(u,\partial_\epsilon u)
\end{align}
Both of these equations can be useful in theoretical and numerical calculations.

Finally, we use Eq.~\eqref{hosusy} with $k=3$ to obtain:
\begin{equation}
	V_3(x)=V_0(x)-2\, \partial_x^2 \ln \left[W(u_1,u_2,u_3)\right], \label{formula3}
\end{equation}
where either Eq.~\eqref{wronsk31} or Eq.~\eqref{wronsk32} should be employed to calculate a new potential from 3-confluent SUSY. Eqs.~\eqref{wronsk31} and \eqref{wronsk32} have 4 parameters, i.e., $\epsilon , C_1, D_1,D_2$. In this case $C_2$ is the constant that disappears due to the choice of $u_1$.

On the other hand, according to Eq.~\eqref{hopsi} the added eigenstate of the new potential $V_3(x)$ with eigenvalue $\epsilon$ is:
\begin{equation}
	\psi_3 (x,\epsilon)=\frac{D_1 + W\left(u,\partial_\epsilon u\right)}{W(u_1,u_2,u_3)}.\label{psi3}
\end{equation}

\section{Differential formula for 4-confluent}\label{fourth}

\subsection{Jordan block of length 4}\label{4jordan}
Although the procedure is clear and we could already generalize it to $k$th-order, we will also analyze explicitly the fourth-order method because we will use it in the applications in Section~\ref{apps}. The Jordan block of length 4 is
	\begin{subequations}
	\begin{align}
		(H-\epsilon)u_1 &=0,\\
		(H-\epsilon)u_2 &=u_1,\\
		(H-\epsilon)u_3 &=u_2,\\
		(H-\epsilon)u_4 &=u_3,
	\end{align}\label{jordan4}
\end{subequations}
\hspace{-1mm}where the solutions $u_1$, $u_2$, and $u_3$ are given by Eqs. \eqref{u1}, \eqref{u2} and \eqref{u3}, while $u_4$ will be calculated next.
\subsection{Eigenvectors of rank 4}\label{4eigen}
Once again, $u_4$ is composed by a homogeneous and an inhomogeneous part. The inhomogeneous part can be calculated from the parametric derivative of Eq.~\eqref{para2}, i.e.,
\begin{equation}
	(H-\epsilon)\frac{\partial_\epsilon^3 u}{3}=\partial_\epsilon^2 u,\label{para3}
\end{equation}
and from here we obtain $u_4^{(p)}$ and then the general solution $u_4$ as
\begin{equation}
	u_4=C_3u+D_3v+C_2\partial_\epsilon u +D_2\partial_\epsilon v+\frac{C_1}{2}\partial_\epsilon^2 u+\frac{D_1}{2}\partial_\epsilon^2 v+\frac{\partial_\epsilon^3 u}{2\cdot 3},
	\label{u4}
\end{equation}
where $C_3, D_3\in \mathbb{R}$ are new constants.

\subsection{Formula for 4-confluent SUSY}\label{4formula}
Now we calculate the Wronskian of $u_1,u_2,u_3,u_4$ as given by Eqs. \eqref{u1}, \eqref{u2}, \eqref{u3} and \eqref{u4}, to obtain (it contains $4!$ terms):
\begin{align}\label{wronsk41}
	W(u_1,u_2,u_3,u_4)=&(C_1D_1D_2-D_1^2C_2-D_2^2+D_1D_3)W(u,v,\partial_\epsilon u,\partial_\epsilon v)\nonumber\\
	&+\left(C_1^2D_1-\frac{C_2D_1^2}{2}-C_1D_2+\frac{D_3}{2}\right)W(u,v,\partial_\epsilon u,\partial_\epsilon^2 u)\nonumber\\
	&+\left(C_1D_1^2-D_1D_2\right)W(u,v,\partial_\epsilon u,\partial_\epsilon^2 v)
	+\left(\frac{C_1D_1-D_2}{6}\right)W(u,v,\partial_\epsilon u,\partial_\epsilon^3 u)\nonumber\\
	&+\left(C_1D_1^2-\frac{D_1D_2}{2}\right)W(u,v,\partial_\epsilon v,\partial_\epsilon^2 u)
	+D_1^3W(u,v,\partial_\epsilon v,\partial_\epsilon^2 v)
	+\frac{D_1^2}{6}W(u,v,\partial_\epsilon v,\partial_\epsilon^3 u)\nonumber\\
	& +\frac{D_1^2}{2}W(u,v,\partial_\epsilon^2 u,\partial_\epsilon^2 v)
	+\frac{D_1}{12}W(u,v,\partial_\epsilon^2 u,\partial_\epsilon^3 u)
	+\left(C_1D_1-\frac{D_2}{2}\right)W(u,\partial_\epsilon u,\partial_\epsilon v,\partial_\epsilon^2 u)\nonumber\\
	& +D_1^2W(u, \partial_\epsilon u,\partial_\epsilon v,\partial_\epsilon^2 v)
	+\frac{D_1}{6}W(u, \partial_\epsilon u,\partial_\epsilon v,\partial_\epsilon^3 u)
	+\frac{D_1}{2}W(u, \partial_\epsilon u,\partial_\epsilon^2 u,\partial_\epsilon^2 v)\nonumber\\
	&+\frac{1}{12}W(u, \partial_\epsilon u,\partial_\epsilon^2 u,\partial_\epsilon^3 u).
\end{align}
Using again the Wronskian identities from the Appendix, in particular Eq.~\eqref{app4} to reduce the order of the Wronskian, we get
\begin{equation}
W(u_1,u_2,u_3,u_4)=W(u_1,u_2)[W(u_1,u_4)+W(u_2,u_3)]-W^2(u_1,u_3), \label{wronsk42}
\end{equation}
where each of these terms is given by:
\begin{subequations}
	\begin{align}
		W(u_1,u_2) =\, & D_1+ W(u,\partial_\epsilon u),\\
		W(u_1,u_4)+W(u_2,u_3)=\, &\, C_1D_2-C_2 D_1 +D_3+C_1^2W(u,\partial_\epsilon u) +3C_1D_1W(u, \partial_\epsilon v)+\frac{3}{2}C_1W(u,\partial_\epsilon^2 u)\nonumber\\
		& +\frac{3}{2}D_1 W(u, \partial_\epsilon^2 v) +\frac{1}{6}W(u, \partial_\epsilon^3 u) 
		+ D_1^2 W(v, \partial_\epsilon v) +\frac{1}{2}W(\partial_\epsilon u, \partial_\epsilon^2 u),\\
		W(u_1,u_3) =\, & D_2+C_1W(u, \partial_\epsilon u)+D_1 W(u, \partial_\epsilon v)+\frac{1}{2}W(u,\partial_\epsilon^2 u).
	\end{align}\label{wronsk43}
\end{subequations}
\hspace{-1mm}Moreover, from Eq.~\eqref{hosusy} with $k=4$ we obtain
\begin{equation}
	V_4(x)=V_0(x)-2\, \partial_x^2 \ln \left[W(u_1,u_2,u_3,u_4)\right], \label{formula4}
\end{equation}
and using either Eq.~\eqref{wronsk41} or Eqs. \eqref{wronsk42} and \eqref{wronsk43} we can calculate a new potential arising from 4-confluent SUSY. The constant $C_3$ does not appear in Eqs.~\eqref{wronsk41} and \eqref{wronsk42}, as we expected now considering previous cases. Therefore we have a 6-parametric family of new potentials, with $\epsilon , C_1, C_2, D_1, D_2, D_3$.

The new potential $V_4(x)$ will have an extra eigenvector in $\epsilon$, which is given by:
\begin{equation}
	\psi_4 (x,\epsilon)=\frac{W(u_1,u_2,u_3)}{W(u_1,u_2,u_3,u_4)},\label{psi4}
\end{equation}
where $W(u_1,u_2,u_3)$ is given either by Eq.~\eqref{wronsk31} or Eq.~\eqref{wronsk32}.

\section{Differential formula for $k$-confluent}\label{kth}
\subsection{Jordan block of length $k$}\label{kjordan}
We can now foresee the structure of the solutions for a Jordan block of arbitrary length $k$. Next, we will prove our hypothesis using mathematical induction assuming our formulas are valid for $k$ and then proving its validity for the $k+1$ case.

First, we need to obtain the $k$ solutions $u_j$ involved in Eq.~\eqref{hosusy}. We start from a Jordan block of length $k$ for the Hamiltonian $H$ and the factorization energy $\epsilon$ 
\begin{subequations}
	\begin{align}
		(H-\epsilon)u_1 & = 0,\label{schseed}\\
		(H-\epsilon)u_2 & = u_1, \\
		\vdots \nonumber \\
		(H-\epsilon)u_k & = u_{k-1}.
	\end{align}\label{jordank}
\end{subequations}
\hspace{-1mm}The Schr\"odinger equation for the seed solution $u$:
\begin{equation}
	(H-\epsilon)u=0.\label{sch11}
\end{equation}
which is chosen from the general solution from Eq.~\eqref{schseed} as we saw previously. Then, let us obtain the parametric derivative of this equation with respect to $\epsilon$
\begin{equation}
	(H-\epsilon)\partial_\epsilon u=u.
\end{equation}
Deriving again
\begin{equation}
	(H-\epsilon)\frac{\partial_\epsilon^2 u}{2}=\partial_\epsilon u,
\end{equation}
and one more time
\begin{equation}
	(H-\epsilon)\frac{\partial_\epsilon ^3 u}{3}=\partial_\epsilon^2 u.
\end{equation}
Now, we are ready to prove by induction the formula for the $k$th derivative of $u$. Let us start from the induction hypothesis given by
\begin{equation}
	(H-\epsilon)\frac{\partial_\epsilon^k u}{k}=\partial_\epsilon^{k-1}u,
\end{equation}
and apply $\partial_\epsilon$ on both sides to obtain the general formula for the index $k+1$:
\begin{equation}
	(H-\epsilon)\frac{\partial_\epsilon^{k+1} u}{k+1}=\partial_\epsilon^{k}u.\label{parak}
\end{equation}
\hfill $\square$

\subsection{Eigenvectors of rank $k$}\label{keigen}
Now we can solve the Jordan block of length $k$ from Eq.~\eqref{jordank}. From the previous sections, we propose that the system is closed by $u_1=u$ and the following $k-1$ functions $u_j$ ($j=2,\dots , k$):
\begin{equation}
	u_j=\sum_{l=1}^{j-1}C_{j-l}\frac{\partial_{\epsilon}^{l-1}u}{(l-1)!}+\sum_{l=1}^{j-1}D_{j-l}\frac{\partial_{\epsilon}^{l-1}v}{(l-1)!}+
	\frac{\partial_\epsilon^{j-1}u}{(j-1)!},\label{uj}
\end{equation}
where, as usual, $\partial^0_\epsilon f=f$ and $0!=1$. Furthermore, this equation reduces to Eq.~\eqref{u2} for $j=2$, the base case for our induction proof.

Now, let us assume that this equation is valid for the $k$th-order case and let us try to prove it for $k+1$ case. From Eq.~\eqref{uj} for $k+1$, we have
\begin{equation}
	u_{k+1}=\sum_{l=1}^{k}C_{k-l+1}\frac{\partial_{\epsilon}^{l-1}u}{(l-1)!}+\sum_{l=1}^{k}D_{k-l+1}\frac{\partial_{\epsilon}^{l-1}v}{(l-1)!}+\frac{\partial_\epsilon^{k}u}{k!}.\label{indk}
\end{equation}
By applying the operator $(H-\epsilon)$ to this function and using the Schr\"odinger equations for $u$ and $v$ given by Eqs.~\eqref{schu} and \eqref{schv} and the general formula for the parametric derivative in Eq.~\eqref{parak}, we obtain
\begin{equation}
	(H-\epsilon)u_{k+1}=\sum_{l=2}^{k}C_{k-l+1}\frac{\partial_{\epsilon}^{l-2}u}{(l-2)!}+\sum_{l=2}^{k}D_{k-l+1}\frac{\partial_{\epsilon}^{l-2}v}{(l-2)!}+\frac{\partial_\epsilon^{k-1}u}{(k-1)!},\label{indk1}
\end{equation}
where we make use of the fact that for $l=1$ we obtain Eqs.~\eqref{schu} and \eqref{schv}. Then changing the labels $l\rightarrow l+1$ in Eq.~\eqref{indk1} we have
\begin{equation}
	(H-\epsilon)u_{k+1}=\sum_{l=1}^{k-1}C_{k-l}\frac{\partial_{\epsilon}^{l-1}u}{(l-1)!}+\sum_{l=1}^{k-1}D_{k-l}\frac{\partial_{\epsilon}^{l-1}v}{(l-1)!}+\frac{\partial_\epsilon^{k-1}u}{(k-1)!},\label{indk2}
\end{equation}
i.e.,
\begin{equation}
	(H-\epsilon)u_{k+1}=u_k,
\end{equation}
which finishes our proof.\hfill $\square$

\subsection{Formula for $k$-confluent SUSY}\label{kformula}
Using the $k$ solutions from Eq.~\eqref{uj} and inserting them into the Wronskian formula for SUSY QM of Eq.~\eqref{hosusy} we can obtain a $k$-confluent transformation to, in principle, any order $k$. These formulas become increasingly complicated; nevertheless, we must remember that we are performing a higher-order confluent transformation, in which $k$ states converge to the same energy ($k\geq 2$), the corresponding integral equations for the usual method of confluent SUSY QM are more complicated to solve, and for several systems they cannot even be calculated.

Also, it is expected that the $k$-function Wronskian that appears in this $k$th-order transformation can be reduced as we did in the third- and fourth-order cases, because the basis fulfills the Schr\"odinger equations \eqref{schu} and \eqref{schv}, which can be used to lower the order of the derivatives in two, i.e.,
\begin{equation}
	\partial_x^l u_j \rightarrow f_1(\partial_x^{l-2} u_j, \partial_x^{l-3} u_j, \dots ,\partial_x u_j, u_j)
\end{equation}
where $f_1$ is a function to be found. Thus, the final Wronskian of $k$ functions will only be dependent on the $k$ functions themselves and their first derivatives with respect to $x$, as it was the case for the second-, third- and fourth-orders, shown in Eqs. \eqref{wronsk2}, \eqref{wronsk32} and \eqref{wronsk42}, respectively, i.e.,
\begin{equation}
	W(u_1,u_2,\dots ,u_k)=f_2(u_1,u_2,\dots ,u_k,\partial_x u_1,\partial_x u_2,\dots ,\partial_x u_k),
\end{equation}
where $f_2$ is another function to be found. We must note that these $u_j$ functions will ultimately depend on the parametric derivatives with respect to $\epsilon$ of the seed solution $u$ and its orthogonal function $v$ up to ($k-1$)th-order, 
\begin{equation}
	W(u_1,u_2,\dots ,u_k)=f_3(u,\dots ,\partial_\epsilon^{k-1}u,v,\dots ,\partial_\epsilon^{k-1}v,\partial_x u,\dots ,\partial_\epsilon^{k-1}\partial_x u,\partial_x v,\dots ,\partial_\epsilon^{k-1}\partial_x v),
\end{equation}
where $f_3$ is a different function. All the derivatives $\partial_x$ will be due two-function Wronskians in the variable $x$ or
\begin{equation}
	W(u_1,u_2,\dots ,u_k)=f_4[W(\partial_\epsilon^\alpha \{u,v\},\partial_\epsilon^\beta \{u,v\})],
\end{equation}
where $0\leq \alpha,\beta \leq k-1$ are integer indexes, $\{u,v\}$ means either one of the functions $u,v$ and $f_4$ is a function to be found. Furthermore, these Wronskians will define a ($2k-2$)-parametric family of new potentials with parameters $\epsilon ,C_1,\dots , C_{k-2},D_1,\dots ,D_{k-1}$. These are the type of equations shown in Eqs. \eqref{wronsk2}, \eqref{wronsk32} and \eqref{wronsk42}.

Furthermore, the new potential $V_k(x)$ could have an extra eigenstate in $\epsilon$ given by Eq.~\eqref{hopsi}.

\section{Applications}\label{apps}

In this section we will apply the third- and fourth-order formulas that we have obtained in this work to two interesting quantum systems. The first of them is the free particle, where both the differential and integral algorithms for the confluent SUSY transformation can be worked out. This will help us to test our ideas and compare our method with the previously known transformations. The second system is the single-gap Lam\'e potential, for which the integral formulation is not easy to apply. As far as we know, this is the first application of 3- and 4-confluent SUSY to the Lam\'e potential, in either integral or differential Wronskian formalism, because they involve elliptic functions, which are difficult to deal with.

\subsection{Free particle}
A free particle is not subject to any force so we can choose its potential as $V=0$. In order to apply our algorithm we need two functions $u,v$ such that $W(u,v)\neq 0\,\forall\, x\in \mathbb{R}$ and for simplicity we will choose $W(u,v)=1$. These functions represent a basis for the two-dimensional solution space associated with the equation $(H-\epsilon)u=0$. We can choose 
\begin{equation}
	(u,v)=\left(\text{e}^{\kappa x},-\frac{\text{e}^{-\kappa x}}{2\kappa}\right),
\end{equation}
where $\epsilon=-\kappa^2$ and for calculating the derivatives we will use the chain rule as $\partial_\epsilon = (\partial_\epsilon \kappa)\, \partial_\kappa$. The domain of this problem is $x\in (-\infty , \infty)$, i.e., the full real-line.

For the $3$-confluent SUSY transformation, we take $u_1=u$, we compute the other two generalized eigenvectors $u_2,u_3$ and then calculate its Wronskian using either Eq.~\eqref{wronsk31} or Eq.{\eqref{wronsk32}, leading to
\begin{equation}
	W(u_1,u_2,u_3)=\frac{1}{8\kappa^3} [4 D_1^2 \kappa^2\text{e}^{- \kappa x} - 8 \kappa^2 (C_1 D_1 \kappa - D_2 \kappa - D_1 x)\text{e}^{\kappa x} - \text{e}^{3 \kappa x}].
\end{equation}
In addition, using Eq.~\eqref{formula3}, the new potential is given by
\begin{equation}
	V_3(x) = -\frac{64 k^3\text{e}^{2 k x}(\text{e}^{2 k x} - D_1 k)[(1 + C_1 k^2 - k x)D_1\text{e}^{2 k x}-2 D_1D_2 k^3-D_2 k^2\text{e}^{2 k x} + 2 D_1^2 k (-1 + C_1 k^2 - k x)]}
	{[\text{e}^{4 \kappa x} - 4 D_1^2 \kappa^2 + 8\kappa^2\text{e}^{2 k x}(C_1 D_1 \kappa - D_2 \kappa - D_1 x)]^2}. \label{v3free}
\end{equation}
From SUSY QM \cite{FS11}, it is known that the 3-confluent transformation on the free particle potential where only one energy level is involved leads to the P\"oschl-Teller potential, but the expression in Eq.~\eqref{v3free} does not look like the P\"oschl-Teller potential since it has three free parameters ($C_1,D_1,D_2$) instead of one. What happens is that we are performing a general 3-confluent transformation on the free particle and indeed we obtain more freedom to choose these parameters, however if we require that the new potentials should not have any singularities we have to restrict to $D_1=0$ and consequently, this result does not depend on $C_1$. We have:
\begin{equation}
	V_3(x)=\frac{64 D_2 \kappa^5 \text{e}^{2 \kappa x}}{(\text{e}^{2 \kappa x} - 8D_2 \kappa^3)^2},
\end{equation}
and if we reparametrize $D_2$ in terms of $x_2$ as
\begin{equation}
	D_2=-\frac{\text{e}^{-2\kappa x_2}}{8\kappa^3},
\end{equation}
then we obtain
\begin{equation}
	V_3(x)=-2\kappa^2\, \text{sech}^2[\kappa(x+x_2)],
\end{equation}
which is indeed the P\"oschl-Teller potential with two free parameters: the factorization energy $\epsilon=-\kappa^2$ and the translation parameter $x_2$. In Fig. \ref{free3} we can see three examples of potentials and their corresponding added physical states obtained from the 3-confluent SUSY.

Therefore, even though we have obtained a more general family of potentials from the $k$-confluent transformation, if we restrict the results to non-singular potentials we go back to simpler ones that are equivalent to those obtained by a first-order SUSY transformation.

\begin{figure}
	\centering
	\includegraphics[scale=0.37]{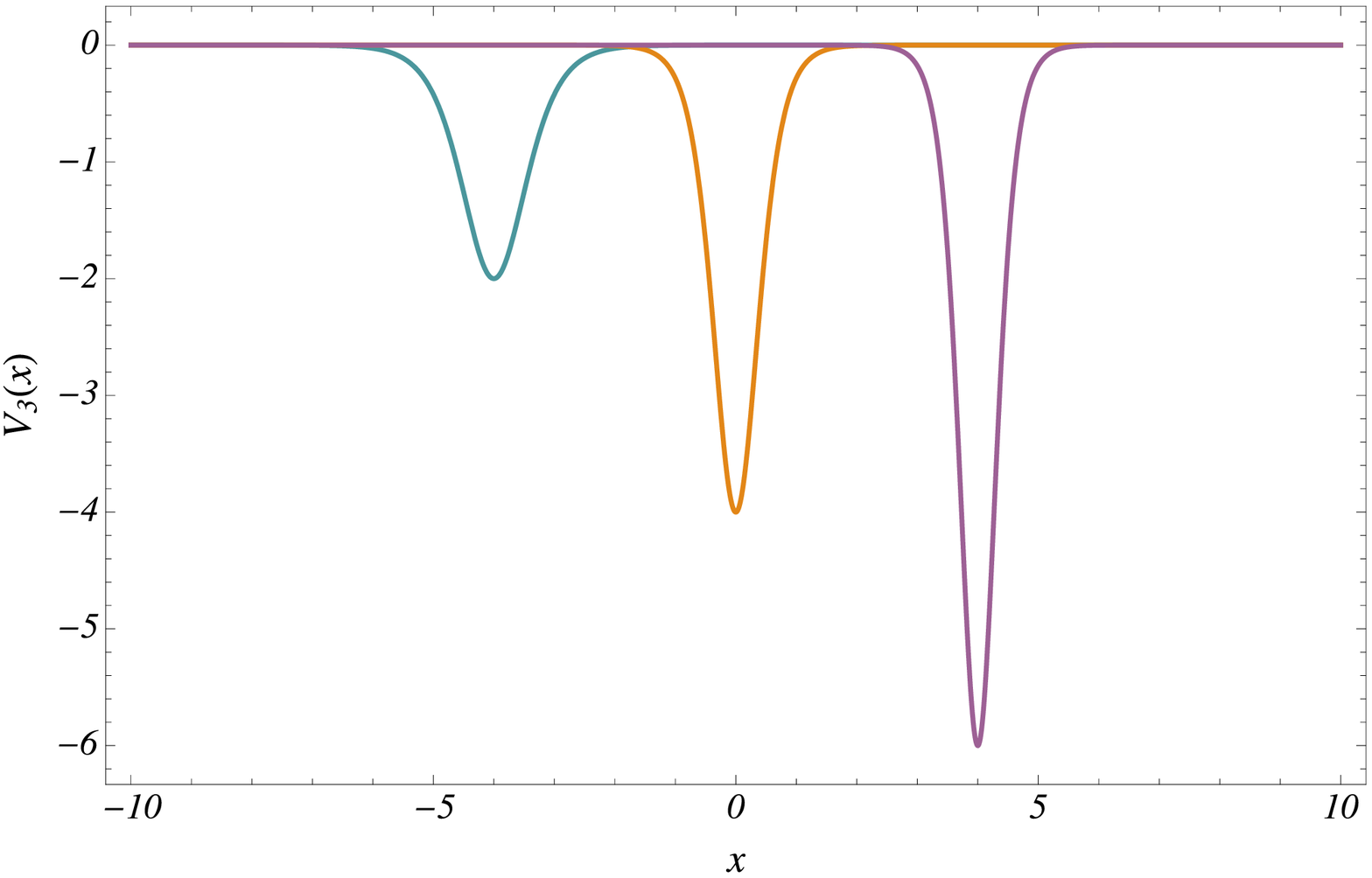}
	\includegraphics[scale=0.37]{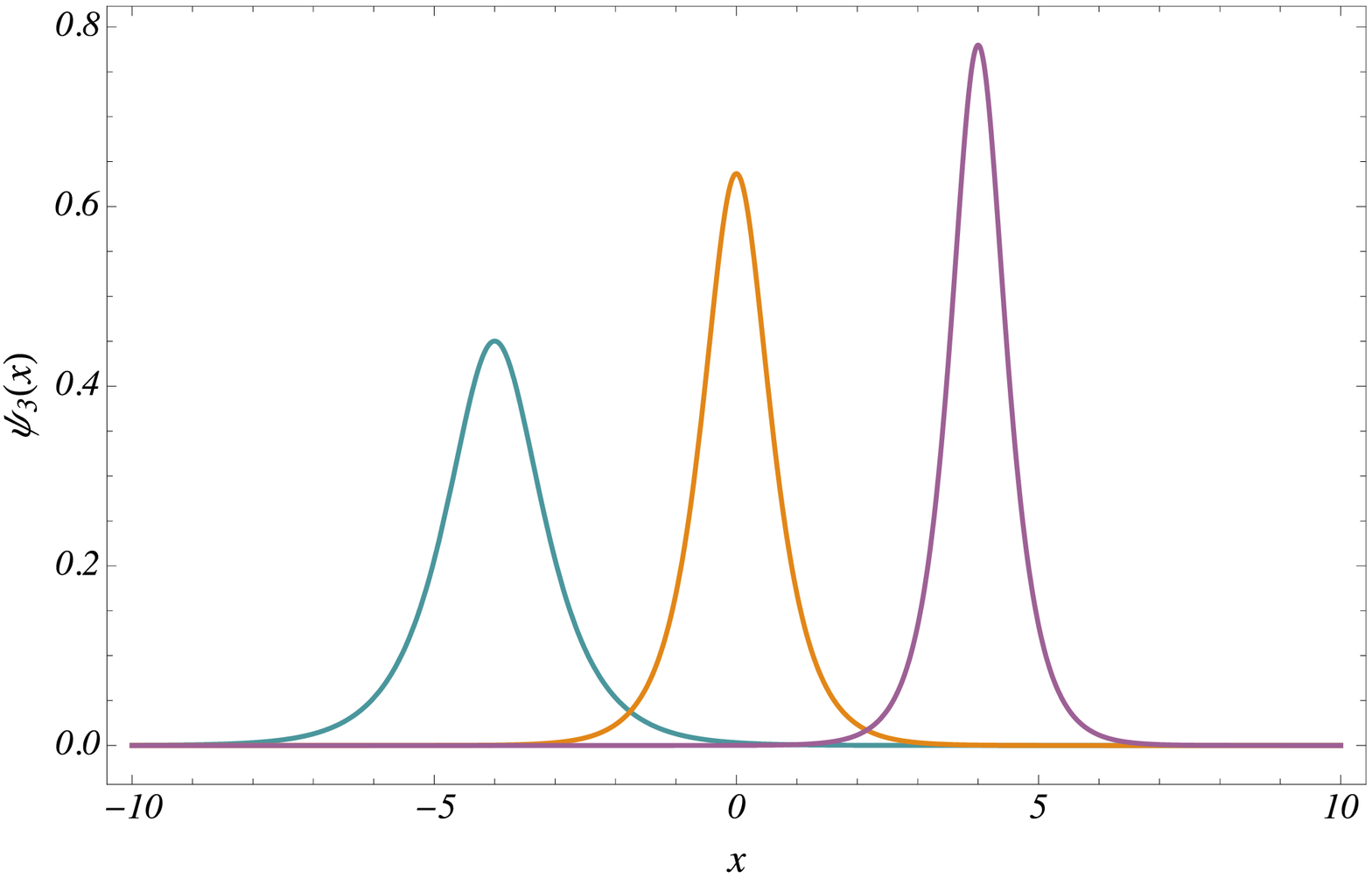}
	\caption{\small{Three P\"oschl-Teller potentials and their corresponding added physical states at $\epsilon$ obtained through 3-confluent SUSY transformations. The plots correspond to $\epsilon=-1, x_2=4$ (blue), $\epsilon=-2, x_2=0$ (orange) and $\epsilon=-3, x_2=-4$ (purple). (Color online).}}
	\label{free3}
\end{figure}

Now, in order to calculate the 4-confluent SUSY transformation we can use either Eq.~\eqref{wronsk41} or Eq.~\eqref{wronsk42}; as both calculations are easy in this case. After some algebra we obtain:
\begin{align}\label{w4free}
	W(u_1,u_2,u_3,u_4)=&
	\frac{\text{e}^{4\kappa x}}{64\kappa^6} +\frac{\text{e}^{2\kappa x}}{16\kappa^5}[ 8\kappa^4(C_1D_2+C_2D_1-C_1^2D_1-D_3) +8\kappa^3 x(C_1D_1-D_2)\nonumber\\
	&  +4\kappa^2(D_2-C_1D_1-D_1x^2)+2D_1\kappa x -D_1]-\frac{D_1^2 x}{4\kappa}+\frac{D_1}{2\kappa^2}\left(C_1 D_1-D_2-D_1 x^2\right)\nonumber\\
	&+\frac{D_1 x}{\kappa}(C_1 D_1-D_2) +C_1 D_1 D_2+D_1 D_3-C_2 D_1^2-D_2^2+\frac{D_1^3 \text{e}^{-2\kappa x}}{8\kappa^3}.
\end{align}

From here, we can calculate the new potential $V_4(x)$; the result is prohibitively long to be written here, but it can be easily calculated using a symbolic computation software. Nevertheless, even though we obtained a 6-parametric family of potentials only a subset of them is non-singular. Furthermore, all of the non-singular sub-families we were able to find can be rewritten as P\"oschl-Teller potentials.

In that case, the P\"oschl-Teller potential should be included in this more general family. In order to prove this, let us take the Wronskian from Eq.~\eqref{w4free} and simplify it by taking the sub-family of potentials with $C_1=D_1=D_2=0$, then
\begin{equation}
	V_4(x)=\frac{256 D_3 k^7 \text{e}^{2 \kappa x}}{(\text{e}^{2\kappa x}-32 D_3\kappa)^2},
\end{equation}
and similarly to the 3-confluent case, we reparametrize $D_3$ in terms of $x_3$ as:
\begin{equation}
	D_3=-\frac{\text{e}^{-2\kappa x_3}}{32\kappa^5},
\end{equation}
thus we can obtain
\begin{equation}
	V_4(x)=-2\kappa^2\, \text{sech}^2[\kappa(x+x_3)],
\end{equation}
which is again the P\"oschl-Teller potential. 

It has been proven that if we apply the 2-confluent SUSY transformation to the free particle we obtain the P\"oschl-Teller potential if we restrict ourselves to the non-singular case (for the differential algorithm see Ref. \cite{BFF12}, for the integral one see Ref. \cite{FS03}). Here we used the differential algorithm we have just developed to calculate the 3- and 4-confluent SUSY transformations from the free particle and we also obtain P\"oschl-Teller potentials for the non-singular case with both methods (see Fig.~\ref{free3}). It is known that a non-singular SUSY transformation for the free particle that adds a single new level to this system will always lead to a P\"oschl-Teller potential, a one-soliton KdV potential \cite{FS05,BFF12,AGP13,MS91}, and this is the case for the $k$-confluent transformation, so our result agrees with the previous ones. Nevertheless, it does not have to be this way for all potentials, in fact, the outcome for the single-gap Lam\'e potential in the next Section points out to a different direction.

\subsection{Single-gap Lam\'e potential}
The Lam\'e periodic potentials are given by \cite{Ars81,FMRS02,FMRS02b}:
\begin{equation}
	V(x) =n(n+1) m\, \text{sn}^2(x|m)=n(n+1)\left[\wp(x+\omega')+\frac{1}{3}(m+1)\right],\label{lame}
\end{equation}
where $\text{sn}(x|m)$ is a Jacobi elliptic function and $\wp(x)$ is the Weierstrass elliptic function, both of them are single-valued doubly-periodic functions \cite{AS72,Cha85} with
\begin{equation}
	K(m)=\int_0^{\pi /2} \frac{d\theta}{(1-m\sin^2\theta)^{1/2}},
\end{equation}
and $\omega'=iK(1-m)$ the real and imaginary quarter-periods of the former and half-periods of the latter. The potentials defined by Eq.~\eqref{lame} have $2n+1$ band edges which define $n+1$ allowed and $n+1$ forbidden bands. The Lam\'e potentials have been used to model a non-relativistic electron in periodic electric and magnetic field configurations which produce a 1D crystal \cite{CJNP08}. Furthermore, they are particular cases of the associated Lam\'e potentials that have been studied in the context of higher-order SUSY QM by Fern\'andez et al. \cite{FG07}. In addition, these potentials are described by an exotic SUSY structure with four supercharges and two bosonic generators \cite{AGP13,CJP08,PN10,AP14,ACJG14} and they also display hidden symmetries related to the soliton dynamics \cite{CP07,AS09}.

In this work we shall deal with the simplest case for $n=1$, known as the single-gap Lam\'e potential. The spectrum of the associated Hamiltonian is given by:
\begin{equation}
	\text{Sp}(H) = [m,1] \cup [1+m,\infty),
\end{equation}
i.e., it is composed of one finite energy band $[m,1]$ and one semi-infinite $[1+m,\infty)$ (see the white region in Fig.~\ref{esplame}). The structure of the resolvent set of $H$ is similar, namely, there is a semi-infinite forbidden energy band $(-\infty,m)$ plus a finite one $(1,1+m)$ (observe the dark zone in Fig.~\ref{esplame}). In solid-state physics, this system is very useful to describe the electron dynamics in semi-conductors. The lower allowed energy band is called \textit{valence band} and the upper one is the \textit{conduction band}; also the finite forbidden energy band is called the \textit{band gap} (see Ch. 5 of Ref. \cite{Cal74}).

\begin{figure}
	\begin{center}
		\includegraphics[scale=0.5]{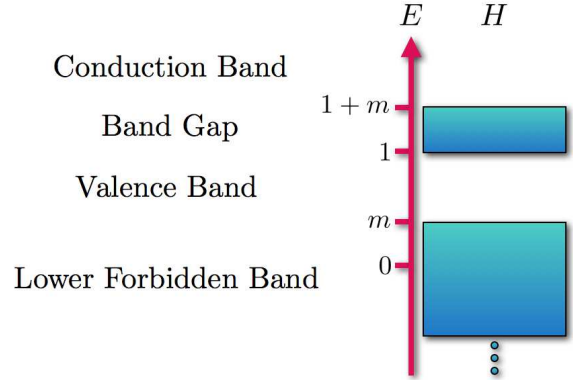}
	\end{center}
	\vspace{-5mm}
	\caption{Spectrum of the Lam\'e potential with $n=1$. The white bands correspond to the allowed energy region, a semi-infinite $[1+m,\infty)$ and a finite one $[m,1]$. The dark region corresponds to the energy gaps, a semi-infinite $(-\infty,m)$ and a finite one $(1,1+m)$.} \label{esplame}
\end{figure}

To implement the confluent second-order SUSY transformation, the appropriate seed solution $u_1$ has to be associated to a factorization energy $\epsilon_1$ which is inside one of the forbidden energy gaps, i.e., the dark zone in Fig.~\ref{esplame}, and such that $W(u_1,u_2) \neq 0 \ \forall \ x\in {\mathbb R}$ all this following the confluent SUSY formalism. Note that in this case, the domain of the eigenvalue problem is also $x\in (-\infty ,\infty )$. For our example this can be achieved by choosing $u_1$ as one of the two Bloch functions associated to $\epsilon$ \cite{FMRS02,FMRS02b}, i.e.,
\begin{align}
	u&=\frac{\sigma(x +\omega ' +\delta)}{\sigma(x+\omega ')}\text{e}^{-x\zeta (\delta)},\\
	v&=-\frac{1}{\sigma^2(\delta) \partial_\delta \wp (\delta)}\frac{\sigma(x +\omega ' -\delta)}{\sigma(x+\omega ')}\text{e}^{x\zeta (\delta)},
\end{align}
where $\sigma$ and $\zeta$ are the non-elliptic Weierstrass functions \cite{Cha85}. Without the constant in $v$, these are the usual (non-normalized) Bloch functions; nevertheless, we choose here a different normalization for both of them, i.e., $u$ without the usual constants and $v$ with a prefactor that depends on $\delta$. This is so for having the minimum necessary factors in order to fulfill the orthogonality condition $W(u,v)=1$, which can be shown using Eqs.~\eqref{identities} and \eqref{sum2} from the Appendix.

Note that $\beta$ is defined by the relation $u(x+T)=\beta u(x)$ and then $\beta=\exp [ 2\delta\zeta(\omega)-2\omega\zeta(\delta) ]$. Besides, by expressing it as $\beta=\text{e}^{i\kappa}$, then $\kappa=2i[\omega\zeta(\delta)-\delta\zeta(\omega)]$ (up to an additive multiple of $2\pi i$) which is known as the quasi-momentum \cite{CJP08}. From the analytical form of this quantity, we can obtain the allowed and forbidden bands by calculating if $\kappa$ takes real or complex values, respectively. The displacement $\delta$ and the factorization energy $\epsilon$ are related by \cite{FMRS02b}:
\begin{equation}
	\epsilon = \frac{2}{3}(m+1)-\wp (\delta). \label{ed}
\end{equation}

It is worth pointing out that we are using one Bloch state to perform the SUSY transformation, even when these states cannot be used for non-confluent SUSY transformations (a linear combination of the two Bloch solutions must be used in that case \cite{FS03,ACJG14}). Nevertheless, one of the advantages of the confluent algorithm is that it can accept a larger set of eigenstates to perform the transformation, for example a state that diverges in one side of the domain. These functions cannot be used in the non-confluent cases and this is exactly the situation for the Lam\'e potential.

We are going to calculate next its parametric derivative with respect to $\epsilon$, for which we will employ some identities for the elliptic functions shown in the Appendix (see Eq.~\eqref{identities}). Then, using the chain rule and Eq.~\eqref{ed} we get:
\begin{equation}
	\partial_\epsilon u = (\partial_\epsilon \delta )\, \partial_\delta u = -(\partial_\delta \wp)^{-1}\partial_\delta u.
\end{equation}
An explicit calculation produces:
\begin{equation}
	\partial_\delta u=[\zeta(x+\delta+\omega')-\zeta(\delta+\omega')+x\wp(\delta)]u,
\end{equation}

With these equations we are now able to calculate the Wronskian using symbolic computation software, either with Eqs. \eqref{wronsk31} or Eqs. \eqref{wronsk32}. In this case, Eq.\eqref{wronsk32} is easier as it is written in terms of simpler Wronskians. The result is too long to be written here explicitly; however, we plot several examples of the potentials and their added states in Fig.~\ref{lame3}. We obtain a 4-parametric family of potential with this algorithm, i.e., $\epsilon ,C_1,D_1,D_2$. As in the previous case, only a subset of these solutions will be non-singular, unfortunately in this case it is not easy to determine a priori all the constraints for the four parameters, although we know that $\epsilon\notin \text{Sp}(H)$.

\begin{figure}
	\centering
	\includegraphics[scale=0.37]{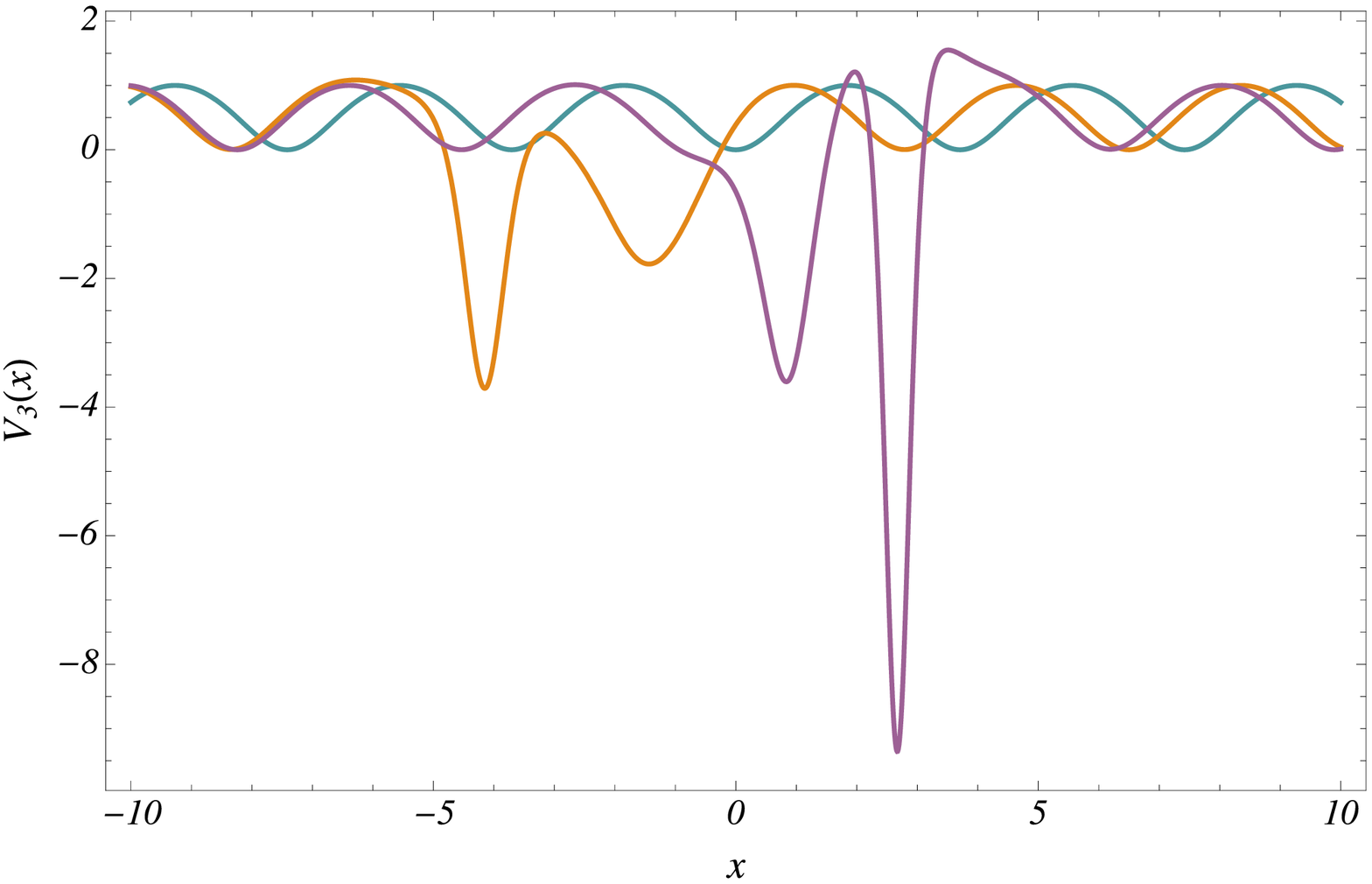}
	\includegraphics[scale=0.37]{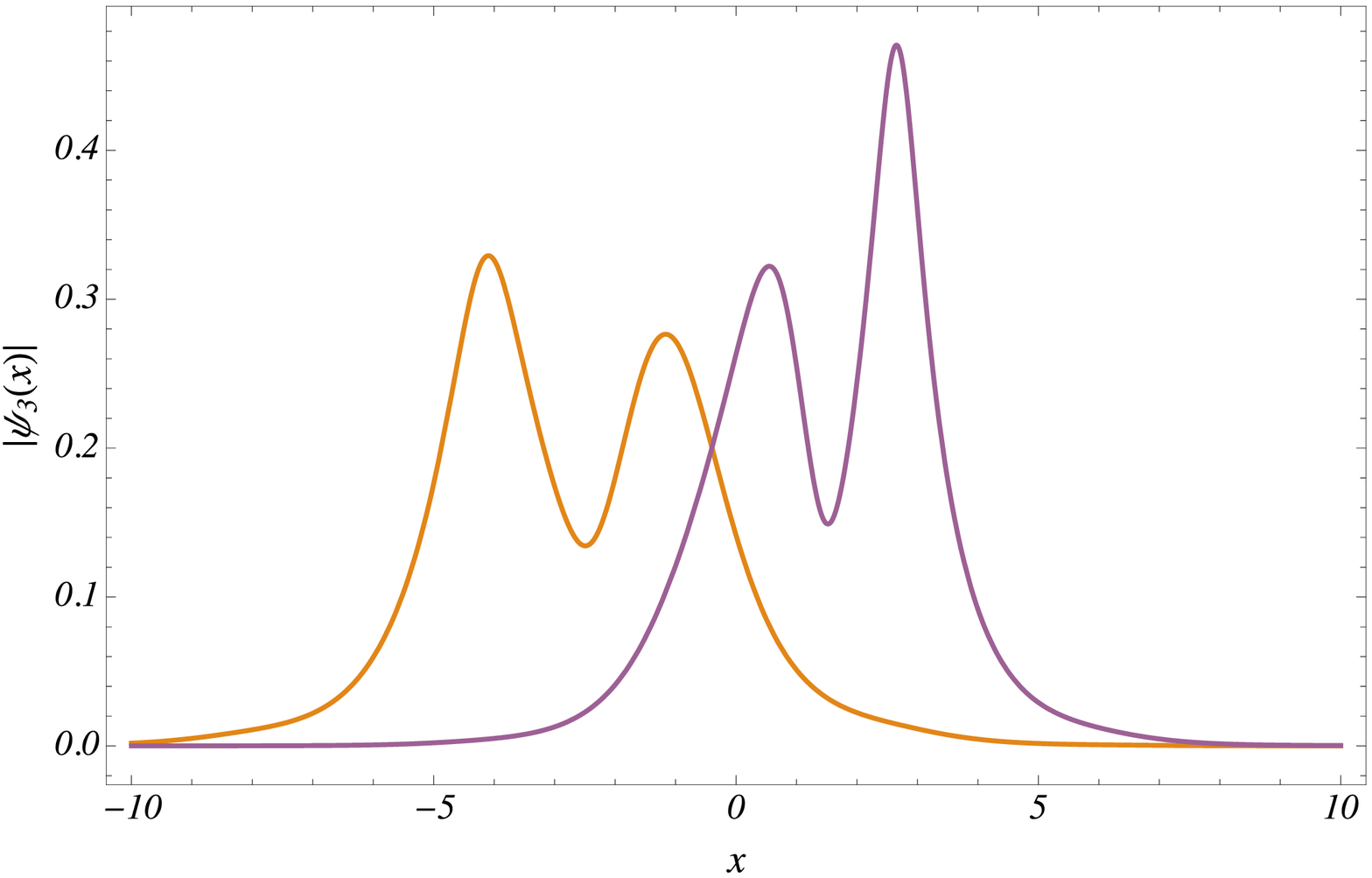}
	\caption{\small{Some SUSY partner potentials and their added states obtained from the single-gap Lam\'e potential using the 3-confluent SUSY transformations. The plots correspond to the original potential $V_0(x)$ with $m=0.5$ (blue) and two modified ones: $V_3(x,0.5,-0.2,0.1,0.01,0.01)$ (orange) and $V_3(x, 0.5, -0.5, 0.1, 10, 0.01)$ (purple) with the format $V_3(x,m,\epsilon,C_1,D_1,D_2)$. As $m=0.5$ and $\epsilon=\{-0.2, -0.5\}$, the transformation energy is in the lower forbidden band. (Color online).}}
	\label{lame3}
\end{figure}

The calculations for the 4-confluent case are much more complicated, but thanks to the fact that this method depends only on calculations of derivatives with respect to $x$ and $\epsilon$, symbolic software is still able to calculate analytically the SUSY partner potentials and the added states for the Lam\'e potentials. Using Eqs. \eqref{wronsk42}, \eqref{wronsk43} and \eqref{formula4} we calculated the new potentials and with Eq.~\eqref{psi4} their physical added states. The results are again too long to be written here but we plot some results in Figs.~\ref{lame4a} and \ref{lame4b}. Given that we are using the confluent algorithm, we can use transformation energy in any of the forbidden bands, including the band gap, which can only be achieved by traditional SUSY transformation by a limiting procedure and leads to the same family of potentials as the confluent case (see Fig. 4 in Ref. \cite{BFF12} and Fig. 8 in Ref. \cite{ACJG14}). We present a a transformed SUSY potential and their added states using transformation functions from the band gap and from the lower forbidden band in Figs.~\ref{lame4a} and \ref{lame4b}, respectively.

We must remark that even though we are only plotting $V_3,\psi_3,V_4,\psi_4$ for the Lam\'e potential, we also have exact analytical expressions for all of them, they are just too long to be written here.

\begin{figure}
	\centering
	\includegraphics[scale=0.37]{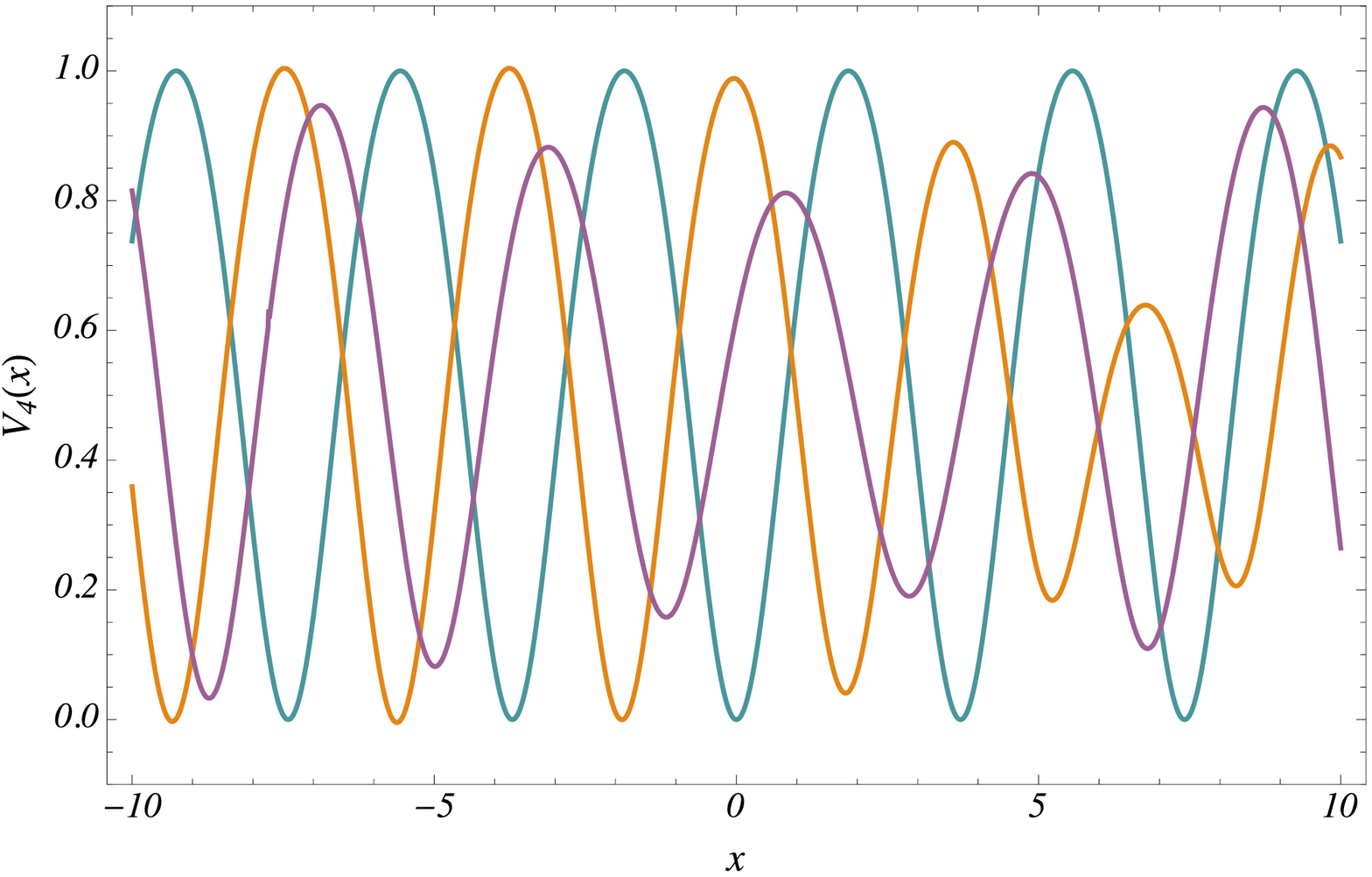}
	\includegraphics[scale=0.37]{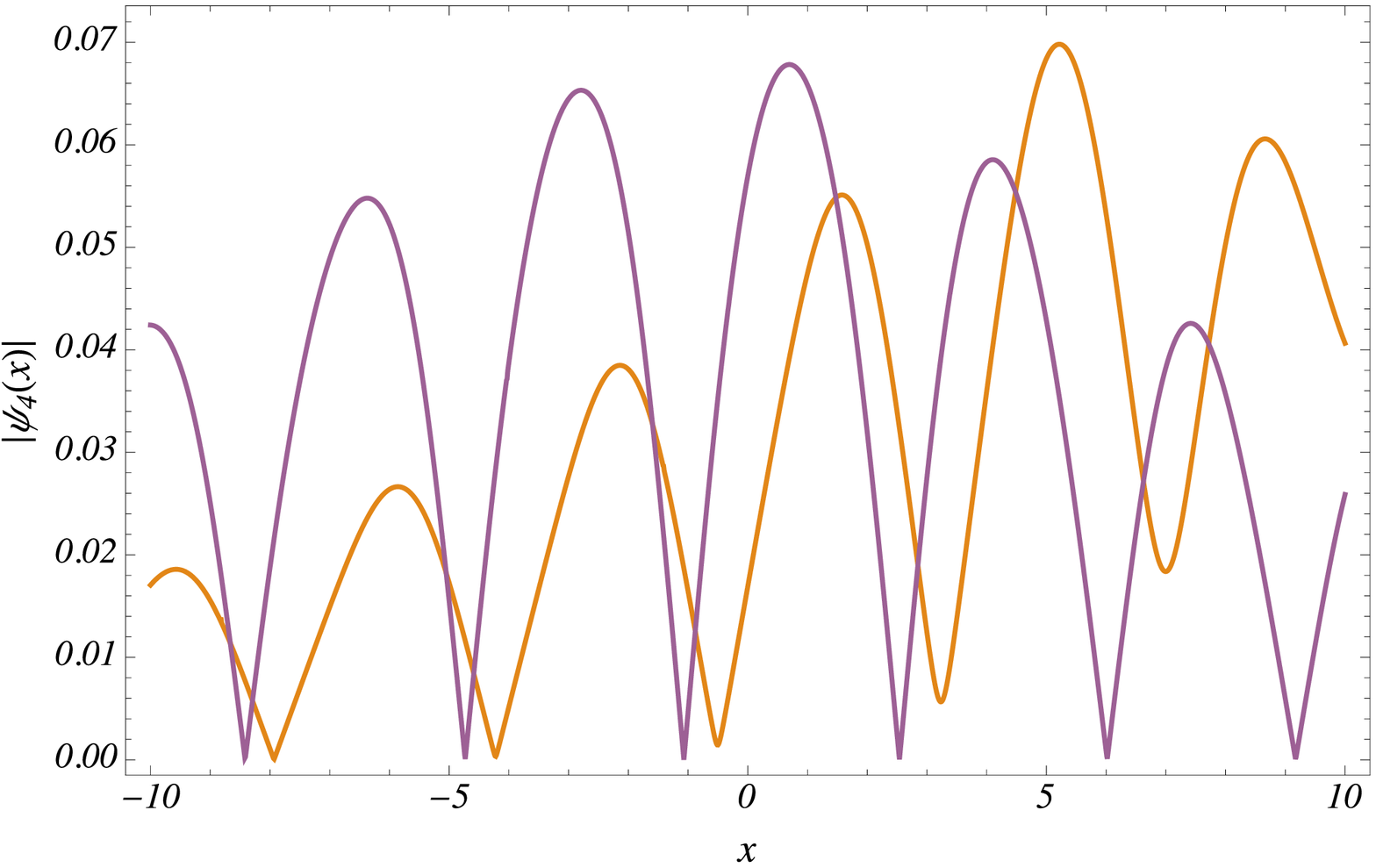}
	\caption{\small{Some SUSY partner potentials and their added states using 4-confluent SUSY transformations departing from the single-gap Lam\'e potential. The plots correspond to the original potential $V_0(x)$ with $m=0.5$ (blue) and two modified ones: $V_4(x,1.25, 0,0,1,0,0,0)$ (orange) and $V_4(x, 1.45, 0,0,0,0,1,0)$ (purple) with the format $V_4(x,m,\epsilon,C_1,C_2,D_1,D_2,D_3)$. In this case the transformation energy belongs to the band gap as $m=0.5$ and $\epsilon=\{1.25,1.45\}$. (Color online).}}
	\label{lame4a}
\end{figure}

\begin{figure}
	\centering
	\includegraphics[scale=0.37]{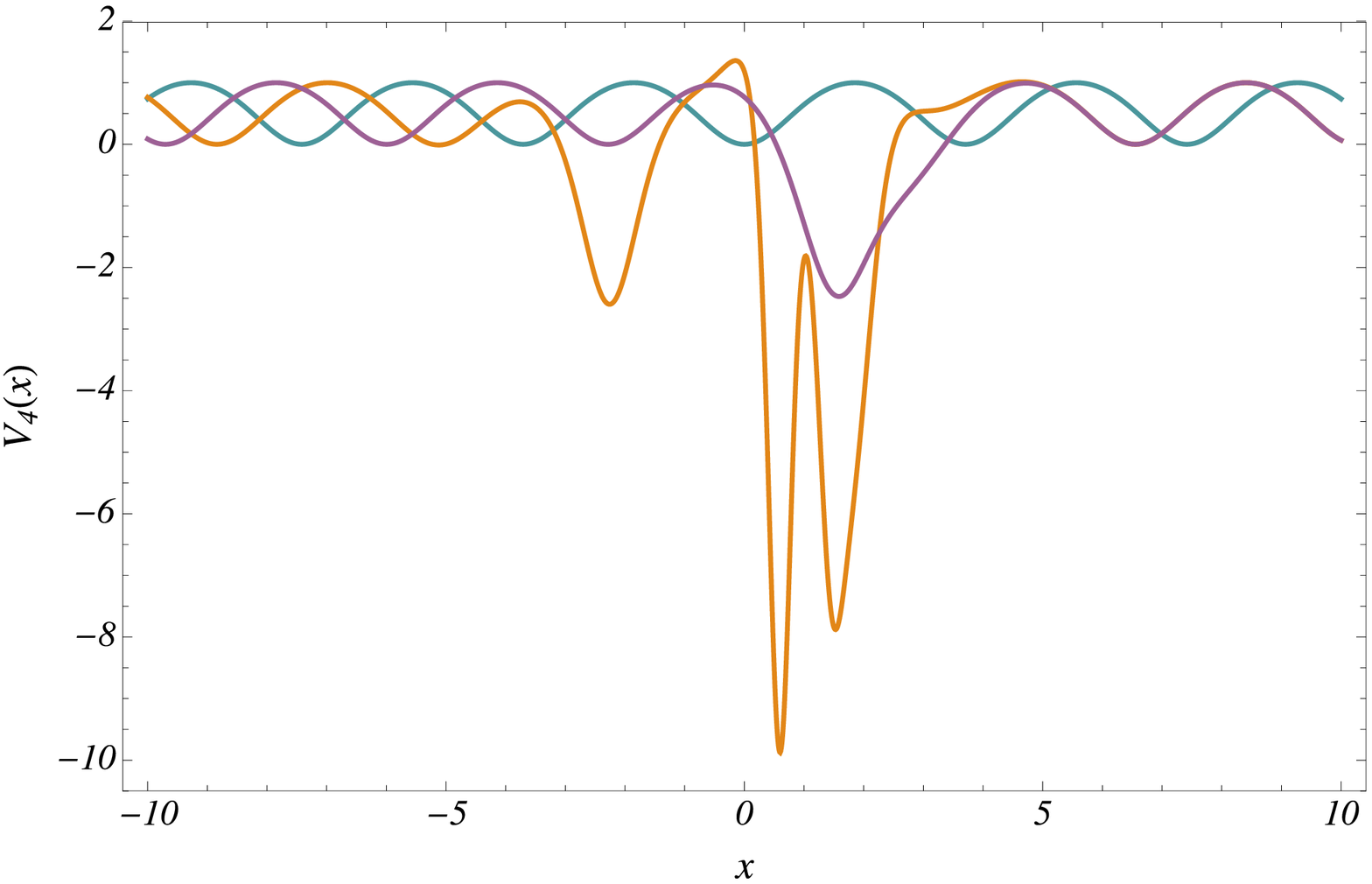}
	\includegraphics[scale=0.37]{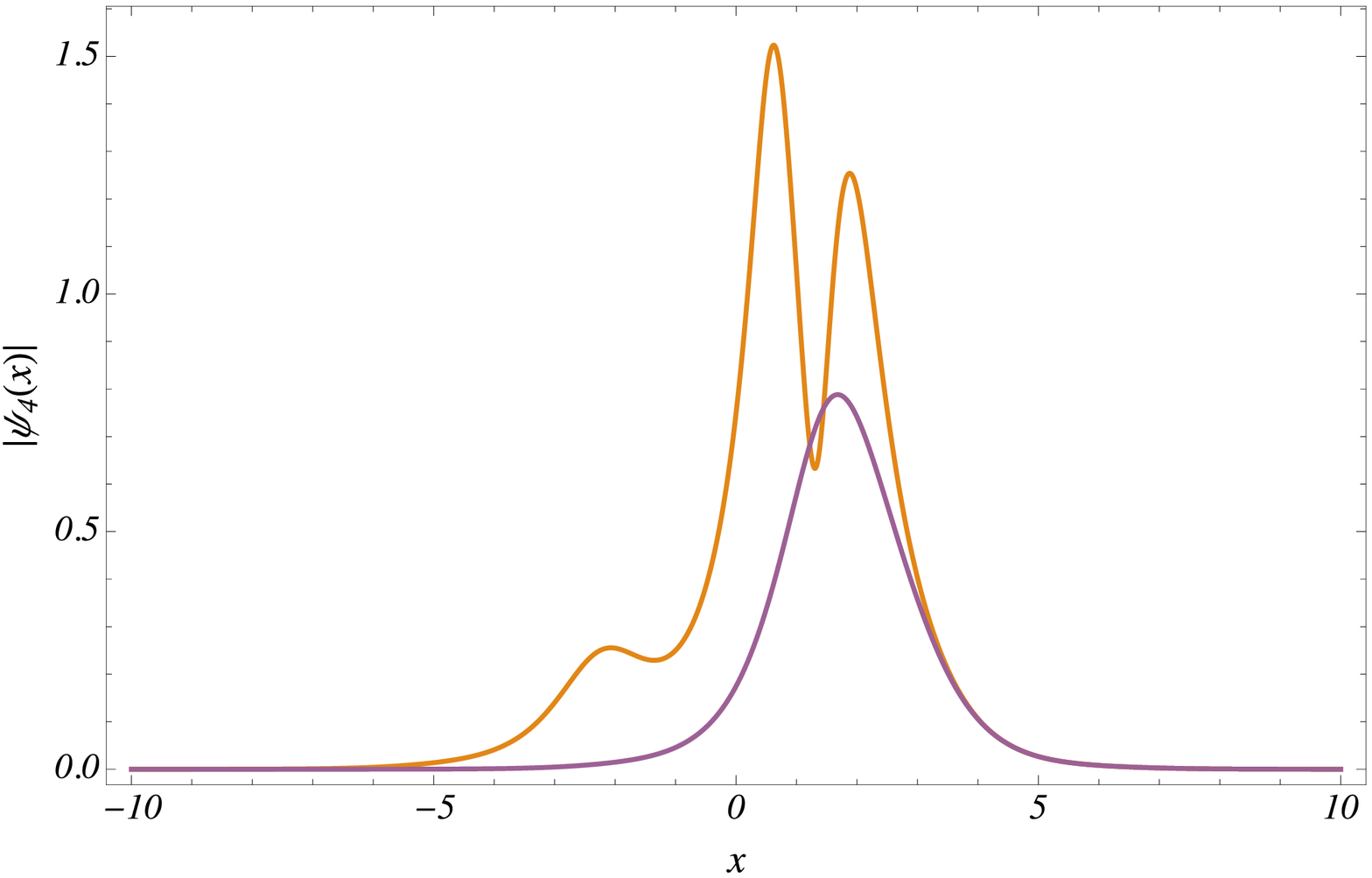}
	\caption{\small{Some SUSY partner potentials and their added states using 4-confluent SUSY transformations departing from the single-gap Lam\'e potential. The plots correspond to the original potential $V_0(x)$ with $m=0.5$ (blue) and two modified ones: $V_4(x,0.5,-1, 1, -2, -1, 2, 3)$ (orange) and $V_4(x, 0.5,-1, 0, 0, 0, 0, -1)$ (purple) with the format $V_4(x,m,\epsilon,C_1,C_2,D_1,D_2,D_3)$. The transformation energy is in the lower forbidden band as $m=0.5$ and $\epsilon=-1$. (Color online).}}
	\label{lame4b}
\end{figure}

\section{Conclusions}
In this work we have developed the general framework for the $k$-confluent SUSY transformation using a Wronskian formula which involves only derivatives with respect to $x$ and to the factorization energy $\epsilon$ instead of integrals, as in the usual $k$-confluent case. In order to do this, we have explicitly worked out the second-, third- and fourth-order cases. Then we have proven the general formulas for the $k$th-order case by mathematical induction in Sec. \ref{kth}.

For the 3- and 4-confluent cases we have also developed some simplifications of the respective Wronskian formulas. We believe that these formulas (see Eqs. \eqref{wronsk31}-\eqref{wronsk32} and \eqref{wronsk41}-\eqref{wronsk43}) can be very useful because they reduce the number of calculations necessary to obtain the transformed potential.

Then, we have applied this formalism to two potentials. The first one is the free particle potential, for which the 3-confluent transformation has been calculated before \cite{FS11}; our calculations here work as a check for our algorithm. We found the expected results already found in the literature. Our formula gives us a 4-parametric family of potentials. Furthermore, we calculated the 4-confluent SUSY partner, which gives us more freedom to deform this potential initially through six parameters, although after we find the conditions to restrict to the non-singular subset, we arrive again to the P\"oschl-Teller potential, exactly the same results as in the 3-confluent case. The reason is that any non-singular potential obtained from the free particle with only one added level will always have this form \cite{FS05,BFF12,AGP13,MS91}.

The second one is the single-gap Lam\'e potential, for which the calculations become much more difficult due to the natural complex nature of the elliptic functions such that most of them are complex and only for special values of their parameters they become real (up to numerical errors in the imaginary part of the order of $10^{-15}$). Nevertheless, we were able to show the possibilities of this method, as for this potential we were able to calculate its 3- and 4-confluent SUSY partners potentials analytically.

We believe this method can be applied to more potentials, as several quantum systems can be solved in terms of the confluent hypergeometric function ${}_1F_1$, and there is a work where their parametric derivatives have been worked out \cite{AG08}.

We must remark the strengths of this method. First, it is a confluent SUSY transformation, which means that it can be applied to a more general class of transformation function than the usual SUSY or Darboux transformation, i.e., using some non-physical states (for the Lam\'e equation we were able to use Bloch states). Second, it is the only way to add only one level above the ground state energy. This case can be obtained as a confluent limit from the usual second-order SUSY transformation. Third, because we have developed the higher-order algorithm and therefore we have now further tools for deforming the potential, because the higher the $k$-confluent transformation, the higher the number of parameters that the new potentials will depend on. Also, we perform this higher-order transformation in a single step. Fourth, the calculation is done through parametric derivatives and Wronskians, unlike the usual confluent case that uses integrals, that most of the times are not easy to solve (they are even unsolvable).

Furthermore, in the case of the Lam\'e potential the possibility of adding a new level in the band gap is significant, because this is the simplest way to model impurities in solid-state physics. The transitions between the valence and the conduction bands are easier when a semi-conductor has impurities, this can be modeled by adding an intermediate physical level in the band gap \cite{Cal74}. An example of this new bound state is $\psi_4$ in the plots of Fig.~\ref{lame4a} and the spectrum will look like the one in Fig.~\ref{addedstate}

\begin{figure}
	\begin{center}
		\includegraphics[scale=0.5]{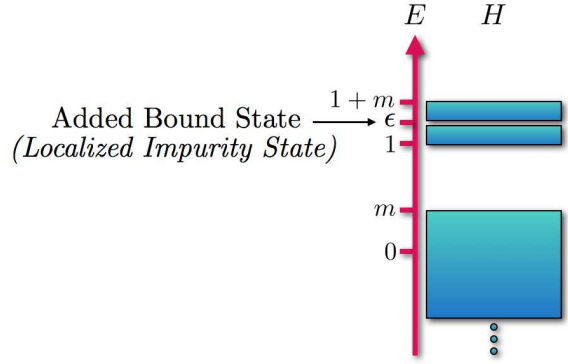}
	\end{center}
	\vspace{-5mm}
	\caption{Spectrum of the deformed single-gap Lam\'e potential. The usual structure of the single-gap Lam\'e potential has one more bound state (the white line in the middle of the band gap), in solid-state physics it is called localized impurity state.} \label{addedstate}
\end{figure}

Finally, there is still perspective work related to the non-singular subfamilies of potentials. Also, it seems likely that there is an the underlying structure for the Wronskian formula for $k$-confluent SUSY which is worth of further study.

\section*{Acknowledgments}

The author would like to thank the anonymous reviewers for their detailed and helpful comments to the manuscript of this work. Also, the author acknowledges the support of Conacyt (Mexico) through Project 152574 and SNI-62547.

\section*{Appendix}
By using the following Wronskian identity
\begin{equation}
	W(f,hf)=(\partial_x h) f^2,\label{wrons}
\end{equation}
which is valid for two differentiable but otherwise arbitrary functions $f$ and $h$, it is straightforward to show that the Wronskian formula for the SUSY transformation given by Eq.~\eqref{hosusy} is preserved for the confluent case.

Using that $u$ and $v$ fulfill the Schr\"odinger equations \eqref{schu} and \eqref{schv} and from the Jordan block of length 3 in Eqs. \eqref{jordan3} after some algebra we get
\begin{equation}
	W(u_1,u_2,u_3)=u_1\, W(u_1,u_3)-u_2\, W(u_1,u_2).\label{app3}
\end{equation}
Similarly, for the Jordan block of length 4 in Eqs. \eqref{jordan4} we get
\begin{equation}
	W(u_1,u_2,u_3,u_4)=W(u_1,u_2)[W(u_1,u_4)+W(u_2,u_3)]-W^2(u_1,u_3).\label{app4}
\end{equation}

Deriving $W(u,v)=1$ with respect to $\epsilon$ we obtain
\begin{equation}
	W(u,\partial_\epsilon v)=W(v,\partial_\epsilon u),
\end{equation}
and deriving again
\begin{equation}
	W(\partial_\epsilon u, \partial_\epsilon v)=\frac{1}{2}[W(u,\partial_\epsilon^2 v)-W(v,\partial_\epsilon^2 u)].
\end{equation}

Also, there are some relationships between $\sigma(x)$, $\zeta(x)$, and $\wp(x)$ and its derivatives that can be found in Ref.~\cite{Cha85} and are shown next:
\begin{subequations}
	\begin{align}
		\partial_x \sigma(x)&=\sigma(x)\zeta(x),\label{sigma}\\
		\partial_x \zeta(x)&=-\wp(x), \label{zeta}\\
		\partial_x \wp(x)&=-\frac{\sigma(2x)}{\sigma^4(x)}.\label{wp}
	\end{align}\label{identities}
\end{subequations}

From Ref.~\cite{AS72} we also have
\begin{subequations}
	\begin{equation}
		\zeta(z_1+z_2)=\zeta(z_1)+\zeta(z_2)+\frac{1}{2}\frac{\partial_{z_1} \wp (z_1)-\partial_{z_2} \wp (z_2)}{\wp (z_1)-\wp (z_2)},
	\end{equation}
	\begin{equation}
		\sigma (z_1+z_2)\sigma (z_1-z_2) =-\sigma(z_1)\sigma(z_2)[\wp(z_1)-\epsilon(z_2)].
	\end{equation}\label{sum2}
\end{subequations}
and the fact that $\wp$ is an even function and $\sigma$ and $\zeta$ are odd.

\bibliographystyle{ieeetr}
\bibliography{references}

\begin{thebibliography}{10}

\bibitem{MR04}
B.~Mielnik and O.~Rosas-Ortiz, ``Factorization: little or great algorithm?,''
  {\em J. Phys. A: Math. Gen.}, vol.~37, pp.~10007--10035, 2004.

\bibitem{Fer10}
D.~Fern\'andez, ``Supersymmetric quantum mechanics,'' {\em AIP Conf. Proc.},
  vol.~1287, pp.~3--36, 2010.

\bibitem{FS03}
D.~Fern\'andez and E.~Salinas-Hern\'andez, ``The confluent algorithm in
  second-order supersymmetric quantum mechanics,'' {\em J. Phys. A}, vol.~36,
  pp.~2537---2543, 2003.

\bibitem{FS05}
D.~Fern\'andez and E.~Salinas-Hern\'andez, ``Wronskian formula for confluent
  second-order supersymmetric quantum mechanics,'' {\em Phys. Lett. A},
  vol.~338, pp.~13--18, 2005.

\bibitem{BFF12}
D.~Bermudez, D.~Fern\'andez, and N.~Fern\'andez-Garc\'ia, ``Wronskian
  differential formula for confluent supersymmetric quantum mechanics,'' {\em
  Phys. Lett. A}, vol.~3756, pp.~692--696, 2012.

\bibitem{Sch13}
A.~Schulze-Halberg, ``Wronskian representation for confluent supersymmetric
  transformation chains of arbitrary order,'' {\em Eur. Phys. J. Plus},
  vol.~128, no.~68, p.~17, 2013.

\bibitem{CS15}
A.~Contreras-Astorga and A.~Schulze-Halberg, ``The generalized zero-mode
  supersymmetry scheme and the confluent algorithm,'' {\em Ann. Phys.},
  vol.~354, pp.~353--364, 2015.

\bibitem{CS15b}
A.~Contreras-Astorga and A.~Schulze-Halberg, ``On integral and differential
  representations of {J}ordan chains and the confluent supersymmetry
  algorithm,'' {\em J. Phys. A: Math. Theor.}, vol.~48, no.~315202, p.~16,
  2015.

\bibitem{GQ15}
Y.~Grandati and C.~Quesne, ``Confluent chains of {DBT}: Enlarged shape
  invariance and new orthogonal polynomials,'' {\em SIGMA}, vol.~61, p.~26,
  2015.

\bibitem{BF13b}
D.~Bermudez and D.~Fern\'andez, ``Factorization method and new potencials from
  the inverted oscillator,'' {\em Ann. Phys.}, vol.~333, pp.~290--306, 2013.

\bibitem{CS14}
A.~Contreras-Astorga and A.~Schulze-Halberg, ``The confluent supersymmetry
  algorithm for {D}irac equations with pseudoscalar potentials,'' {\em J. Math.
  Phys.}, vol.~55, no.~103506, p.~16, 2014.

\bibitem{AP15}
A.~Arancibia and M.~Plyushchay, ``Chiral asymmetry in propagation of soliton
  defects in crystalline backgrounds,'' {\em arXiv:1507.07060 [hep-th]}, 2015.

\bibitem{CJP15}
F.~Correa, V.~Jakubsky, and M.~Plyushchay, ``{PT}-symmetric invisible defects
  and confluent {D}arboux-{C}rum transformations,'' {\em Phys. Rev. A},
  vol.~92, no.~023839, p.~14, 2015.

\bibitem{FH15}
M.~Fiset and V.~Hussin, ``Supersymmetric infinite wells and coherent states,''
  {\em arXiv:1502.05452 [math-ph]}, 2015.

\bibitem{SW15}
A.~Schulze-Halberg and J.~Wang, ``Confluent supersymmetric partners of quantum
  systems emerging from the spheroidal equation,'' {\em Symmetry}, vol.~7,
  pp.~412--426, 2015.

\bibitem{FS11}
D.~Fern\'andez and E.~Salinas-Hern\'andez, ``Hyperconfluent third-order
  supersymmetric quantum mechanics,'' {\em J. Phys. A: Math. Theor.}, vol.~44,
  no.~365302, p.~11, 2011.

\bibitem{Cru55}
M.~Crum, ``Associated {S}turm-{L}iuville systems,'' {\em Quart. J. Math.
  Oxford}, vol.~2, no.~6, pp.~121--127, 1955.

\bibitem{Ber13}
D.~Bermudez, {\em Polynomial {H}eisenberg algebras and {P}ainlev{\'e}
  equations}.
\newblock Ph{D} thesis, Cinvestav, Mexico, July 2013.

\bibitem{DK67}
P.~Dennery and A.~Krzywicki, {\em Mathematics for physicists}.
\newblock New York: Dover Publications, 1967.

\bibitem{HJM06}
E.~Hern\'andez, A.~J\'auregui, and A.~Mondrag\'on, ``Non-{H}ermitian degeneracy
  of two unbound states,'' {\em J. Phys. A: Math. Gen.}, vol.~39,
  no.~10087--10105, 2006.

\bibitem{Mat92}
V.~Matveev, ``Generalized {W}ronskian formula for solutions of the {KdV}
  equations: first applications,'' {\em Phys. Lett. A}, vol.~166, pp.~205--208,
  1992.

\bibitem{Sta95}
A.~Stahlhofen, ``Completely transparent potentials for the {S}chr{\"o}dinger
  equation,'' {\em Phys. Rev. A}, vol.~51, pp.~934--943, 1995.

\bibitem{AGP13}
A.~Arancibia, J.~M. Guilarte, and M.~Plyushchay, ``Effects of scalings and
  translations on the supersymmetric quantum mechanical structure of soliton
  systems,'' {\em Phys. Rev. D}, vol.~87, no.~045009, p.~23, 2013.

\bibitem{MS91}
V.~Matveev and M.~Salle, {\em Darboux transformations and solitons}.
\newblock Berlin: Springer, 1991.

\bibitem{Ars81}
F.~Arscott, {\em Periodic differential equations}.
\newblock Cambridge: Cambridge University Press, 1981.

\bibitem{FMRS02}
D.~Fern\'andez, B.~Mielnik, O.~Rosas-Ortiz, and B.~Samsonov, ``The phenomenon
  of {D}arboux displacements,'' {\em Phys. Lett. A}, vol.~294, p.~168, 2002.

\bibitem{FMRS02b}
D.~Fern\'andez, B.~Mielnik, O.~Rosas-Ortiz, and B.~Samsonov, ``Nonlocal
  supersymmetric deformations of periodic potentials,'' {\em J. Phys. A: Math.
  Gen.}, vol.~35, pp.~4279--4291, 2002.

\bibitem{AS72}
M.~Abramowitz and I.~Stegun, {\em Handbook of mathematical functions with
  formulas, graphs and mathematical tables}.
\newblock New York: Dover Publications, 1972.

\bibitem{Cha85}
K.~Chandrasekharan, {\em Elliptic functions}.
\newblock Berlin: Springer-Verlag, 1985.

\bibitem{CJNP08}
F.~Correa, V.~Jakubsky, L.~Nieto, and M.~Plyushchay, ``Self-isospectrality,
  special supersymmetry, and their effect on the band structure,'' {\em Phys.
  Rev. Lett.}, vol.~101, no.~030403, p.~4, 2008.

\bibitem{FG07}
D.~Fern\'andez and A.~Ganguly, ``Exactly solvable associated {L}am\'e
  potentials and supersymmetric transformations,'' {\em Ann. Phys.}, vol.~332,
  pp.~1143--1161, 2007.

\bibitem{CJP08}
F.~Correa, V.~Jakubsky, and M.~Plyushchay, ``Finite-gap systems,
  tri-supersymmetry and self-isospectrality,'' {\em J. Phys. A: Math. Theor.},
  vol.~41, no.~485303, p.~35, 2008.

\bibitem{PN10}
M.~Plyushchay and L.~Nieto, ``Self-isospectrality, mirror symmetry, and exotic
  nonlinear supersymmetry,'' {\em Phys. Rev. D}, vol.~82, no.~065022, p.~12,
  2010.

\bibitem{AP14}
A.~Arancibia and M.~Plyushchay, ``Transmutations of supersymmetry through
  soliton scattering and self-consistent condensates,'' {\em Phys. Rev. D},
  vol.~90, no.~025008, p.~15, 2014.

\bibitem{ACJG14}
A.~Arancibia, F.~Correa, V.~Jakubsk\'y, J.~Guilarte, and M.~Plyushchay,
  ``Soliton defects in one-gap periodic system and exotic supersymmetry,'' {\em
  Phys. Rev. D}, vol.~90, no.~125041, p.~29, 2014.

\bibitem{CP07}
F.~Correa and M.~Plyushchay, ``Hidden supersymmetry in quantum bosonic
  systems,'' {\em Ann. Phys.}, vol.~322, pp.~2493--2500, 2007.

\bibitem{AS09}
A.~Andrianov and A.~Sokolov, ``Hidden symmetry from supersymmetry in
  one-dimensional quantum mechanics,'' {\em SIGMA}, vol.~5, no.~64, p.~26,
  2009.

\bibitem{Cal74}
J.~Callaway, {\em Quantum theory of the solid state}.
\newblock New York: Academic Press, 1974.

\bibitem{AG08}
L.~Ancarani and G.~Gasaneo, ``Derivatives of any order of the confluent
  hypergeometric function {${}_1F_1(a,b,z)$} with respect to the parameter
  {$a$} or {$b$},'' {\em J. Math. Phys.}, vol.~49, no.~063508, p.~16, 2008.

\end{thebibliography}
\end{document}